\documentclass{ws-ijmpa}
\usepackage{graphicx}

\begin{document}
\newcommand{\met}{{\ensuremath{\not\!\!E_T}}}

\title{Radiative decay $Z_H\to \gamma A_H$ in the little Higgs
model with T-parity}
\author{I. Cort\'es-Maldonado and G. Tavares--Velasco}
\address{Facultad de Ciencias
F\'\i sico Matem\'aticas, Benem\'erita Universidad Aut\'onoma de
Puebla, Apartado Postal 1152, Puebla, Pue., M\'exico}

\date{\today}
 \maketitle
\begin{abstract}
In the little Higgs model with T-parity (LHTM), the only tree-level kinematically
allowed two-body decay of the $Z_H$ boson can be the $Z_H\to A_H H$ decay and thus
one-loop induced two-body
decays may have a significant rate. We study the  $Z_H\to \gamma A_H$ decay, which is induced at the
one-loop level by a fermion triangle and is interesting as it depends on the mechanism of
anomaly cancellation of the model. All the relevant two- and three-body decays of the
$Z_H$ gauge boson arising at the tree-level are also calculated. We considered a small
region of the parameter space where $f$ is still allowed to be as low as $500$ GeV  by
electroweak precision data. We analyzed the scenario of a Higgs
with a mass of 120 GeV. We found that the $Z_H\to\gamma
A_H$ branching ratio can
be of the order of a tree-level three-body decay and may be at the reach of
detection at the
LHC for $f$ close to 500 GeV, but it may be difficult to detect for $f=1$ TeV. There is
also an scenario where the Higgs boson has an intermediate mass such that the $Z_H\to A_H
H$ decay is closed, the $Z_H\to \gamma A_H$ gets considerably enhanced and the chances of
detection get a large boost.

\end{abstract}

\ccode{PACS number(s):14.70.Pw,13.38.Dg}

\section{Introduction}
\label{int}

Little Higgs models offer a solution to the little hierarchy problem based on the idea that the
Higgs boson is a pseudo-Goldstone boson arising from an approximately broken global symmetry
associated with a strongly interacting sector. Although some models based on this idea were already
proposed long ago,\cite{Kaplan:1983fs,Kaplan:1983sm} they were unsuccessful as there was
the need to reintroduce  fine-tuning to obtain a light Higgs boson. A solution to this
problem was proposed in Ref.
\refcite{ArkaniHamed:2001nc}: by invoking a collective mechanism of symmetry breaking,
the gauge and Yukawa couplings of the Goldstone boson are introduced in such a way that the Higgs
boson mass is free of quadratic divergences at the one-loop  or even at the two-loop level.
Several realizations of this idea have been proposed in the literature, but the most popular is the
littlest Higgs model (LHM).\cite{ArkaniHamed:2002qy} Apart from reproducing the SM at
the electroweak scale, the LHM predicts heavy partners for the SM gauge bosons and
the top quark, which are necessary to cancel the quadratic divergences of the Higgs boson mass at
the one-loop level.  However, this model predicts large corrections to electroweak precision
observables and the scale of the global symmetry breaking, $f$, is constrained by experimental
data to be larger than about $4000$ GeV.\cite{Csaki:2002qg} One alternative to evade this
strong constraint relies on the introduction of a discrete symmetry called
T-parity,\cite{Low:2004xc} which forbids any
dangerous contributions to electroweak observables and allows for much weaker constraints
on $f$.\cite{Hubisz:2005tx} The littlest Higgs model with T parity (LHTM) has become the
source of considerable interest in the literature recently.
\cite{Hubisz:2004ft,Belyaev:2006jh,Freitas:2006vy}

The LHTM is a nonlinear sigma model that has a global symmetry under the group $SU(5)$ and a local
symmetry under the subgroup $[SU(2)\times U(1)]^2$. There is two extra neutral gauge bosons,
$Z_H$ and $A_H$, that are the partners of the $Z$ gauge boson
and the photon, respectively. While the $Z_H$ gauge boson is associated with the extra
$SU(2)$ gauge group, the photon partner is associated with the extra $U(1)$
gauge group. The latter particle is the lightest one of the model and is a promising
dark matter candidate. It is not possible to obtain a
model-independent bound on the mass of an extra neutral gauge boson from experimental measurements,
but electroweak precision data\cite{Erler:1999nx} (EWPD) along with
Tevatron\cite{:2005dia} and
LEP2\cite{Alcaraz:2006mx} searches, allow one to obtain limits on its mass from about
$500$ GeV to
$1000$ GeV in models with universal flavor gauge couplings.   While an extra neutral gauge boson
with a mass around 1 TeV may be detected at the LHC,  the future international linear collider would
be able to produce it with a mass up to 2-5 TeV.\cite{Langacker:2008yv} This would open
up potential opportunities to study the phenomenology of this particle.

In the LHTM, the new gauge bosons are T-odd and the SM particles are T-even. Therefore
T-parity invariance imposes severe restrictions on the decay modes of the new particles. While the
heavy photon is stable, the only tree-level kinematically allowed two body-decay of the $Z_H$ gauge
boson is $Z_H\to A_H H$. We are interested in studying the one-loop induced decay
$Z_H\to \gamma A_H$, which may have a significant branching ratio similar to that of a tree-level
three-body decay. This decay is interesting as its signature at particle colliders would be very
peculiar. In addition, the respective decay width could be useful to explore the mechanism of
anomaly cancellation present in the model. Decays of an extra neutral gauge boson into a pair of
neutral gauge bosons have already been studied in the literature, for instance in the context of a
superstring-inspired $E_6$ model,\cite{Chang:1988fq} the minimal 331 model,
\cite{Perez:2004jc} 5D warped-space models,\cite{Perelstein:2010yd} and little Higgs
model without T-parity.\cite{CortesMaldonado:2011pi} The remainder of this paper is
structured as follows. In Section II we
present a survey of the LHTM, with particular emphasis on the gauge sector and the
properties of the extra neutral gauge boson $Z_H$. Section III is devoted to present the calculation
of the one-loop decay  $Z_H\to \gamma A_H$.  We will also
discuss the dominant decay modes of the $Z_H$ boson  arising at the tree-level.
Section IV is devoted to discuss the results, and the conclusions are presented in Sec. V.

\section{The framework of the little Higgs model with T-parity}
\label{model}

In the LHM, the largest
corrections to electroweak precision observables  arise from the heavy gauge
bosons.\cite{Csaki:2002qg}  A global fit  to experimental
data yields a strong constraint on the symmetry breaking scale, $f>4$ TeV, for a wide region of
the space of parameters.\cite{Csaki:2002qg} This would require reintroducing  fine-tuning
to have a light Higgs boson.
Once T-parity is introduced into the model,\cite{Low:2004xc} the tree-level
contributions to electroweak observables arising from the heavy
gauge bosons cancel, and the resulting constraints on the scale $f$ are
significantly weaker: there is an area in the parameter space where $f$ is allowed to be
as low as 500 GeV,\cite{Hubisz:2005tx} which depends on the Higgs boson mass value and
the ratio between the masses of the T-odd and T-even top partners.

\subsection{The scalar and gauge sectors}
The LHM  is a  nonlinear sigma model  with a global symmetry under the $SU(5)$ group and a gauged subroup $[SU(2) \otimes U(1)]^2$. The Goldstone bosons are parametrized by the following $\Sigma$ field
\begin{equation}
\Sigma = e^{i\Pi/f}\ \Sigma_0\  e^{i\Pi^T/f},
\label{sigmaA}
\end{equation}
where $\Pi$ is the pion matrix. The $\Sigma$ field transforms under the gauge group as $\Sigma \to \Sigma' = U\  \Sigma \ U^T$,
with  $U=L_1 Y_1 L_2 Y_2$ an element of the gauge group.

The $SU(5)$ global symmetry is broken down to
$SO(5)$ by the sigma field VEV, $\Sigma_0$, which is of the order of the scale of the symmetry breaking. After the global symmetry is broken, 14 Goldstone bosons arise accommodated in multiplets of the electroweak gauge group: a real singlet, a real triplet, a complex triplet and a complex doublet. The latter will be identified with the SM Higgs doublet. At this stage, the gauge symmetry is also broken to its diagonal subgroup,  $SU(2) \times U(1)$.  The real singlet and the real triplet are absorbed by the gauge bosons associated with the broken gauge symmetry.

The LHM effective Lagrangian is assembled by the kinetic
energy Lagrangian of the $\Sigma$ field, $\mathcal{L}_{\rm K}$,
the Yukawa Lagrangian, $\mathcal{L}_{\rm Y}$, and the kinetic terms of the gauge and fermion sectors.  The sigma field kinetic Lagrangian is given by
\cite{ArkaniHamed:2002qy}
\begin{equation}
\mathcal{L}_{\rm K}= \frac{f^{2}}{8} {\rm Tr} | D_{\mu} \Sigma |^2,
\label{kinlag}
\end{equation}
with the $[SU(2)\times U(1)]^2$ covariant derivative defined by\cite{ArkaniHamed:2002qy}
\begin{equation}
D_{\mu} \Sigma = \partial_{\mu} \Sigma
- i \sum_{j=1}^2 \left[ g_{j} W_{j\,\mu}^{a} (Q_{j}^{a}\Sigma + \Sigma Q_{j}^{a\,T})
+ g'_{j} B_{j\,\mu} (Y_{j} \Sigma+\Sigma Y_{j}^{T}) \right].
\end{equation}
The heavy $SU(2)$ and $U(1)$ gauge bosons are  $W_{j}^\mu =
\sum_{a=1}^{3} W_{j}^{\mu \, a} Q_{j}^{a}$ and $B_{j}^\mu =
B_{j}^{\mu} Y_{j}$, with  $Q_j^a$ and $Y_j$ the gauge generators,
while $g_i$ and $g'_i$ are the respective gauge couplings. The VEV
$\Sigma_0$ generates masses for the gauge bosons and mixing between them. The heavy gauge boson mass
eigenstates are given
by\cite{ArkaniHamed:2002qy}
\begin{eqnarray}
W'^a &=& -c W_{1}^a + s W_{2}^a,\\
B' &=& -c^{\prime} B_{1} + s' B_{2}  ,
\end{eqnarray}
with masses  $m_{W'} = \frac{f}{2}\sqrt{g_1^2+g_2^2}$ and
$m_{B'} = \frac{f}{\sqrt{20}} \sqrt{g'^2_{1}+g'^2_{2}}$.

The orthogonal combinations of gauge bosons are identified with the SM
gauge bosons:\cite{ArkaniHamed:2002qy}
\begin{eqnarray}
W^a &=& s W_{1}^a + c W_{2}^a,\\
B &=& s' B_{1} + c' B_{2},
\end{eqnarray}
which remain massless at this stage,
their couplings being given by
$g = g_{1}s = g_{2}c$ and
$g' = g'_{1}s' = g'_{2}c'$,
where $s = g_{2}/\sqrt{g_{1}^{2}+g_{2}^{2}}$ and
$s' = g'_{2}/\sqrt{g'^2_{1}+g'^2_{2}}$ are mixing parameters.

The gauge and Yukawa interactions that break the global $SO(5)$ symmetry induce radiatively a Coleman-Weinberg potential, $V_{CW}$, whose explicit form can be obtained after expanding the $\Sigma$ field:
\begin{equation}
    V_{\rm CW} = \lambda_{\phi^2} f^2 {\rm Tr}|\phi|^2
    + i \lambda_{h \phi h} f \left( h \phi^\dagger h^T
        - h^* \phi h^\dagger \right)
    - \mu^2 |h|^2
    + \lambda_{h^4}  |h|^4,
\end{equation}
where $\lambda_{\phi^2}$, $\lambda_{h \phi h}$, and $\lambda_{h^4}$
depend on the fundamental parameters of the model, whereas $\mu^2$, which receives logarithmic divergent
contributions at one-loop level and quadratically divergent
contributions at the two-loop level, is treated as a free parameter
of the order of $f^2/16 \pi^2$. The Coleman-Weinberg potential induces a mass term for the complex
triplet $\Phi$, whose components acquire masses of the order of $f$. The neutral component of the
complex doublet develops a VEV, $v$, of the order of the electroweak scale, which is responsible for
EWSB. The VEV $v$ along with the triplet VEV, $v'$, are obtained when $V_{CW}$ is minimized.

In the gauge sector, T-parity
only exchanges the $[SU(2) \times U(1)]_1$
and $[SU(2) \times U(1)]_2$ gauge bosons: $W_{1}^{\mu \, a} \leftrightarrow W_{2}^{\mu \, a}$ and
$B_{1}^{\mu}\leftrightarrow B_{2}^{\mu}$. T-parity invariance is achieved by setting the coupling
constants at the values $g_1 = g_2$ and $g'_1 =g'_2$.\cite{Low:2004xc} The light SM gauge
bosons
are T-even, while the heavy gauge bosons are T-odd.
At the electroweak scale, EWSB proceeds as usual, yielding the final mass eigenstates: the three SM gauge bosons are accompanied by three heavy gauge bosons which are their counterpart,  $A_H$, $W_H$ and
$Z_H$. The masses of the heavy gauge bosons  get
corrected by terms of the order of $(v/f)^2$ and so are the masses
of the weak gauge bosons $W_L$ and $Z_L$. The heavy gauge boson masses are given by:

\begin{equation}
m_{Z_H}\simeq m_{W_H}=gf \left(1-\frac{v^2}{8f^2}\right),
\label{WHmasslimit}\\
\end{equation}
\begin{equation}
m_{A_H}= \frac{g'f}{\sqrt{5}}\left(1-\frac{5v^2}{8f^2}\right).
\label{AHmasslimit}
\end{equation}

As far as the scalar sector of the theory is
concerned, due to the transformation property of the $\Sigma$ field under T-parity ($\Sigma\to
\bar{\Sigma}=\Sigma_0\Omega \Sigma^\dagger \Omega \Sigma_0$, with $\Omega={\rm diag}(1,1,-1,1,1)$)
the SM Higgs doublet turns out to be T-even, while the additional $SU(2)_L$ triplet $\Phi$ is T-odd.
The $H\Phi H$ coupling is thus forbidden and so is a nonzero  $SU(2)_L$ triplet VEV, $v'$.
After diagonalizing the Higgs mass matrix, the light Higgs boson
mass can be obtained at the leading order
\begin{equation}
    m^2_{H}= 2 \mu^2 = 2 \left( \lambda_{h^4}
    - \frac{\lambda_{h \phi h}^2}{ \lambda_{\phi^2}} \right) v^2
\end{equation}

It is required that  $\lambda_{h^4} > \lambda_{h \phi h}^2 /
\lambda_{\phi^2}$ to obtain the correct  electroweak symmetry
breaking vacuum with $m^2_H>0$. The Higgs triplet masses are
degenerate at this order: $ m_{\Phi}=\sqrt{2} m_H\,f/v$.

In
summary, in the gauge and scalar sectors both the LHM and the LHTM have the same particle
content.\cite{Low:2004xc}  T-parity invariance has important phenomenological
consequences  as there are
only vertices containing an even number of T-odd
particles. The heavy photon, which is the lightest new particle, is stable and thus a  dark
matter candidate. The $Z_H$ gauge boson can only decay into a kinematically allowed odd number of
heavy photons accompanied by SM particles.

\subsection{Fermion sector}

In order to avoid compositeness constraints,\cite{Low:2004xc} T-parity requires two
$SU(2)$ doublets, $q_1$ and $q_2$, for each SM doublet. Under a T-parity
transformation, these doublets are exchanged $q_1\leftrightarrow - q_2$. The T-even (T-odd)
combination of $q_1$ and $q_2$ is the SM (T-odd) fermion doublet. The mass of each T-odd fermion
doublet is generated by the interaction

\begin{equation}
 {\cal L}_\kappa=-\kappa_f\left(\bar{\Psi}_2\xi \Psi_c+\bar{\Psi}_1\Omega \xi^\dagger \Omega \Psi_c\right),
\end{equation}
where the $SU(5)$ multiplets $\Psi_i$ are defined by $\Psi_1=(q_1,0,{\bf 0}_2)^T$ and $\Psi_2=({\bf 0}_2,0,q_2)^T$, with $q_{1,2}=-\sigma_2({u_{1,2}}_L,{d_{1,2}}_L)^T$. Also, the multiplet $\Psi_c=(q_c,\chi_c,\tilde{q}_c)^T$ is introduced such that it transforms nonlinearly under $SU(5)$.  It can be shown that ${\cal L}_\kappa$ is T-parity invariant as the following transformation rules are obeyed: $\Psi_{1,2}\to -\Sigma_0 \Psi_{2,1}$, $\Psi_c\to-\Psi_c$, $\xi\to \Omega \xi^\dagger\Omega$, and $\Sigma\to \Sigma_0 \Omega \Sigma^\dagger \Omega \Sigma_0$.

It is easy to see that the components of the T-odd doublet $q_-=(q_1+q_2)/\sqrt{2}=(i {d_-}_L,-i{u_-}_L)^T$ have the following masses

\begin{eqnarray}
m_{u_-}&\sim& \sqrt{2}\kappa f \left(1-\frac{v^2}{8f^2}\right),\\
m_{d_-}&=&\sqrt{2}\kappa f.
\label{Toddmass}
\end{eqnarray}

The effects of the heavy T-odd fermions has been investigated in Ref.
\refcite{Belyaev:2006jh} and it was
shown that they may be non-negligible at high-energy colliders. For simplicity, an universal value
for $\kappa$ will be assumed for all the T-odd fermions. Also, flavor nondiagonal interactions will
be neglected.

In order to cancel the quadratic divergences to the Higgs boson mass arising from the SM top quark, the top sector must be  additionally modified. The corresponding $SU(5)$ multiplets are completed by introducing two $SU(2)$ singlets ${U_1}_L$ and ${U_2}_L$: $Q_1=(q_1,{U_1}_L,{\bf 0}_2)^T$ and $Q_2=({\bf 0}_2,{U_2}_L,q_2)^T$. The T-parity invariant Yukawa Lagrangian for the top sector can be written as

\begin{eqnarray}
{\cal L}^Y_t&=&\frac{\lambda_1 f}{2\sqrt{2}}
\epsilon_{ijk} \epsilon_{xy} \left((\bar{Q}_1)_i
\Sigma_{jx}\Sigma_{ky}-(\bar{Q}_2\Sigma_0)_i \tilde{\Sigma}_{jx}\tilde{\Sigma}_{ky}\right){u_+}_R\nonumber\\&-&
 \lambda_{2} f \left({\bar{U_1}}_L{U_1}_R+ {\bar{U_2}}_L {U_2}_R  \right)+{\rm H.c.},
\end{eqnarray}
From here we can obtain the mass eigenstates: there are a new T-odd quark
$T_-=({U_1}_L+{U_2}_R)/\sqrt{2}$ and a new  T-even quark $T_+$. The latter together with de top
quark are given by:

\begin{equation}
 \left(\begin{array}{c}
{u_+}_X\\{U_+}_X\end{array}\right)=\left(\begin{array}{cc}
c_X&s_X\\-s_X&c_X\end{array}\right)
\left(\begin{array}{c}
{t_+}_X\\{T_+}_X\end{array}\right),
\end{equation}
for $X=L,R$ and with the T-even eigenstates defined as $u_+=(u_1-u_2)/\sqrt{2}$. The mixing angles are $s_L\sim s_\alpha^2 v/f$ and $s_R\sim s_\alpha=\lambda_1/\sqrt{\lambda_1^2+\lambda_2^2}$. The masses of the new T-odd and T-even  quarks are given to the lowest order by \cite{Hubisz:2004ft}:

\begin{eqnarray}
    m_{T_+} &=& \sqrt{\lambda_1^2 + \lambda_2^2} f,\\
    m_{T_-} &=& \lambda_2f,
\label{Toppartmass}
\end{eqnarray}
with the top mass given by
\begin{equation}
    m_t = \frac{\lambda_1 \lambda_2}{\sqrt{\lambda_1^2 + \lambda_2^2}} v,\label{Topmass}
\end{equation}
The above Yukawa interaction also corrects the SM
couplings with terms of the order of $(v/f)^2$.

In summary, each SM fermion has associated a T-odd fermion with a mass given by Eq. (\ref{Toddmass})
and there is a new T-even top partner $T_+$ that has its associated T-odd fermion $T_-$. The
interactions of the T-even fermions and their T-odd partners with the neutral gauge bosons are given
by:\cite{Belyaev:2006jh}
\begin{equation}
 {\cal L}=\sum_{u,d}\bar{f}_L\gamma^\mu {f_-}_L\left(\left(g c_H {T_3}_f+g' c_H
Y'\right){Z_H}_\mu+\left(-g s_H {T_3}_f+g' c_H Y'\right){A_H}_\mu\right)+{\rm H.c.}
\end{equation}
where $Y'=-1/10$, $s_H\simeq gg'/(g^2-g'^2/5)v^2/(4f^2)$ describes the
degree of mixing between neutral heavy gauge bosons, and $c_H^2=1-s_H^2$. $s_H$. The
corresponding Feynman rules are shown in \ref{Couplings} along with those for the
couplings
of the $Z$ gauge boson to T-odd fermions and also the interaction vertices for the heavy gauge
bosons.

\section{One-loop decay $Z_H\to \gamma A_H$}
\label{cal}

Because of T-parity invariance,  the $Z_H$ gauge boson  can only decay into
one heavy photon plus other SM particles, although the decay into three heavy photons can
also be kinematically allowed. As long as the Higgs boson mass is light,
the only kinematically allowed $Z_H$ tree-level two-body decay is $Z_H\to A_H H$.  The
respective decay width can be
written as

\begin{eqnarray}
\Gamma(Z_H \rightarrow A_H H)&=& \frac{g_{Z_H A_H H}^2} {192 \pi m_{Z_H}y_{A_H}}
\sqrt{(1-(\sqrt{y_H}-\sqrt{y_{A_H}})^2)(1-(\sqrt{y_H}+\sqrt{y_{A_H}})^2)}
\nonumber\\&\times&\left(1+(y_H-y_{A_H})^2+y_Z^2-2(y_H-5y_{A_H})\right).
\label{ZHtoAHH}
\end{eqnarray}
where $g_{Z_H A_H
H}=gg'v/2$ and  $y_{a}=(m_{a}/m_{Z_H})^2$. The following three-body decays are also
allowed: $Z_H\to A_H W W$, $Z_H\to A_H Z Z$, $Z_H\to A_H\bar{f}f$, and even $Z_H\to A_H H H$ and
$Z_H\to 3A_H$. To obtain the
respective decay widths, we squared the decay amplitude with the aid of the FeynCalc
package\cite{Mertig:1990an}
and the integration over the three-body phase space was performed numerically via  the
internal Mathematica routines. We refrain from  presenting the analytical results as they are too
cumbersome to be included here.

\begin{figure}[!ht]
 \centering
\includegraphics*[width=3.5in]{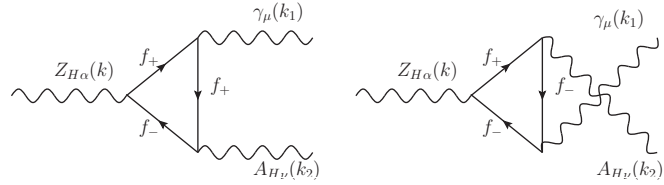}
 \caption{\label{FeynDiag}One-loop Feynman diagrams contributing to the $Z_H$ gauge boson
decay
 $Z_H\to \gamma A_H$ in the LHTM. $f_+$  stands for a T-even charged fermion and $f_-$ for
its associated T-odd charged fermion.}
 \end{figure}

At the one-loop level the decay  $Z_H\to \gamma A_H$ proceeds through a fermion triangle and its
amplitude depends on the mechanism of anomaly cancellation. We expect
that the branching ratio for this decay can compete with those of the tree-level
three-body decays, which are suppressed due to phase-space. The decay $Z_H\to \gamma A_H$ proceeds
via the Feynman diagram of Fig. \ref{FeynDiag},  which involves two fermions of opposite T-parity
circulating in the loop.

The decay amplitude for the $Z_H\to \gamma A_H$ decay can be written as

\begin{eqnarray}
{\cal M}({Z_H}\to \gamma A_H)&=&\frac{1}{m_{Z_H}^2}
\epsilon_{\alpha}(k)\epsilon_{\mu}(k_1)\epsilon_{\nu}(k_2)\Big\{i A_1^{\gamma A_H}
\left(k_1^\nu \epsilon^{\alpha\mu\lambda\rho}+{k_1}^\alpha
\epsilon^{\mu\nu\lambda\rho}\right){k_1}_\lambda {k}_\rho\nonumber\\&+&i A_2^{\gamma A_H}
k_1 \cdot k_2 \epsilon^{\alpha\mu\nu\lambda}{k_1}_\lambda+A_3^{\gamma
A_H}{k_1}^\alpha\left(k_1 \cdot k_2 \, g^{\mu\nu}-{k_1}^\nu{k_2}^\mu \right)
\nonumber\\&+&A_4^{\gamma A_H}{k_1}^\nu\left(k_1 \cdot k_2\,
g^{\alpha\mu}-{k_1}^\alpha{k_2}^\mu\right) \Big\},
\label{MVtoAAH}
\end{eqnarray}
which was arranged in this peculiar form to display electromagnetic gauge invariance. Here
$k_1^\mu$ and $k_2^\nu$ are the 4-momenta of the outgoing $\gamma$ and $A_H$ gauge bosons,
whereas
$k^\alpha$ is the 4-momentum of the $Z_H$ gauge boson. The mass-shell and transversality
conditions, along with Schouten's identity, were used to eliminate redundant terms. The
explicit
form of the $A_i^{\gamma A_H}$ coefficients is shown in \ref{Coefficients} in
terms of Passarino-Veltman scalar functions. Once the $Z_H\to \gamma A_H$ amplitude is
squared and summed (averaged) over polarizations of the ingoing (outgoing) gauge bosons,
the decay width can be written as
\begin{eqnarray}
\Gamma(Z_H\to \gamma A_H)&=&\frac{1}{3}\frac{\left(1-y_{A_H}\right)^5
\left(1+y_{A_H}\right) m_{Z_H}}{2^5 \pi y_{A_H}}\Big(
 |A_1^{\gamma A_H}-A_2^{\gamma A_H}|^2+|A_3^{\gamma A_H}|^2\nonumber\\&+&|A_4^{\gamma
A_H}|^2\Big).
\label{VtoAAH}
\end{eqnarray}

We will evaluate the $Z_H\to \gamma A_H$ branching ratio for values of the parameters
consistent with the constraints from EWPD.

\section{Numerical results and discussion}

\subsection{Current constraints on the LHTM parameter space}
Before presenting our results we would like to discuss  on the current constraints on
the parameter space of the LHTM from EWPD. In Ref.
\refcite{Hubisz:2005tx}, the symmetry breaking scale $f$ was constrained via the oblique
parameters $S$, $T$ and $U$, together with the $Z\to \bar{b}b$ decay.
A similar analysis was done more recently in Ref. \refcite{Baak:2011ze}. It was found
that the allowed values
of the scale $f$ depend on the Higgs boson mass and the ratio of the masses of the T-odd
and T-even top partners, $s_\lambda=m_{T_-}/m_{T_+}$. The contribution to the
$T$ parameter from the each T-odd fermion doublet was found to be \cite{Hubisz:2005tx}

\begin{equation}
T_{{\rm T-odd}}=-\frac{\kappa^2}{192 \pi^2\alpha}\left(\frac{v}{f}\right)^2, 
\end{equation}
Thus the contribution from the T-odd fermions is
negligible as long as they are relatively light, but it can be significant if they
are too heavy. Along these, lines, an upper bound on the T-odd fermion masses can be found
from the LEP
bound on four-fermion interactions:
\cite{Hubisz:2005tx}
\begin{equation}
\label{upbound}
m_{f_-}< 4.8\left(\frac{f}{1 \,\,{\rm TeV}}\right)^2\quad {\rm TeV}.
\end{equation}
This leads to a maximal contribution to
the $T$ parameter of about $0.05$ for
each T-odd fermion as long as its mass reaches this upper
bound.\cite{Hubisz:2005tx} If such a maximal contribution is taken into
account, the allowed area on the $f$ vs $m_H$ plane shrinks
considerably, although $f$ below 1 TeV is still allowed for some values of $s_\lambda$
and $m_H$. \cite{Baak:2011ze}

As far as the direct search of T-odd quarks is concerned, the D0 Collaboration has
obtained a lower bound on the T-odd quark mass from the search for final events with
jets and large missing transverse energy at the Tevatron.\cite{Abazov:2008at}
According to this bound, which depends on the mass of the heavy photon $m_{A_H}$, a light
T-odd quark with a mass of about $100$ GeV is not ruled out by Tevatron data as long as
$m_{A_H}\simeq 100$ GeV. More recently, the search for final events with jets and large
missing transverse energy at the LHC has been used by the CMS \cite{Khachatryan:2011tk}
and Atlas \cite{daCosta:2011qk} Collaborations to search for supersymmetry. An analysis
presented in Ref. \refcite{Perelstein:2011ds} shows that the LHC data can also be used to
impose a bound on the T-odd quark masses that is stronger than the one found
at the Tevatron. It was concluded that $m_{q_-}$ below 450 GeV is ruled out for $m_{A_H}
\simeq 100$ GeV with 95 \% C.L. Furthermore,  the data
collected at the LHC during the years 2011 and 2012 will be useful to place a bound on the
T-quark mass of about 650 GeV for the $m_{A_H}\simeq 300$ GeV and below.cite{Perelstein:2011ds}

\begin{figure}[!ht]
 \centering
\includegraphics[width=3.5in]{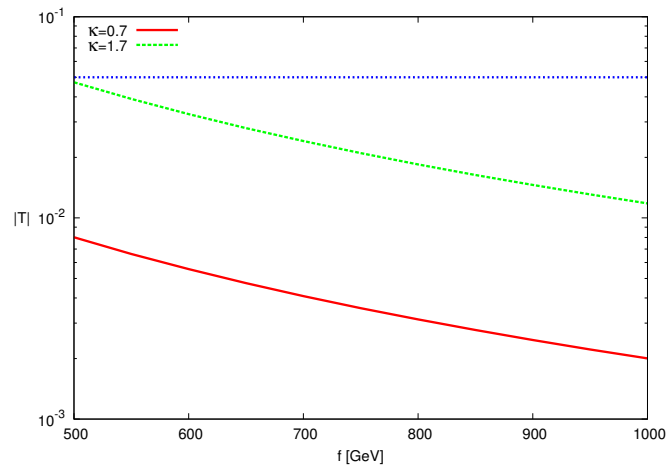}
\caption{\label{Tparam} Contribution of each T-odd fermion doublet to the oblique
parameter $T$ for two values of $\kappa$.  The horizontal line is the maximal
contribution from a T-odd fermion assuming that its mass reaches the
the upper bound from the LEP bound on four fermion contact interactions.}
 \end{figure}

We will see below that  the possible detection of the $Z_H\to \gamma A_H$ decay at the LHC
would be very difficult for $f$ above 1 TeV as the estimated production of $pp\to W_H Z_H$
events would require a branching ratio above the 0.1 level to have just a handful of
$Z_H\to \gamma A_H$ events. We thus must seek for regions of the parameter space where $f$
is still allowed to be below 1
TeV. From the results presented in Ref. \refcite{Baak:2011ze} for the allowed area on the
$f$ vs $m_H$
plane with 95 \% C.L., we can conclude that there are two promising scenarios for
observing the $Z_H\to \gamma A_H$ decay channel at the LHC: one scenario in which the
T-odd fermions are relatively light and another scenario in which they are very heavy. We
will illustrate these scenarios assuming particular values for the parameters $s_\lambda$
and $\kappa$. The first scenario to be analyzed corresponds
to $s_\lambda=0.55$ and the presence of T-odd fermions heavy enough to give a large
contribution
to the $T$ parameter. This scenario  would allow for values of
$f$ as low as about $600$ GeV for a wide range of values of $m_H$. Another potential
scenario
arises when $s_\lambda=0.75$ and the T-odd fermions are
light enough to allow one to
neglect their contribution to the $T$ parameter, thereby allowing a large region of the
parameter space.  To accomplish these two scenarios we can choose
convenient values of
the parameter $\kappa$ to tune the mass of the T-odd fermions. For instance, 
we show in Fig. \ref{Tparam} the contribution to the $T$ parameter from a T-odd
fermion when $\kappa=0.7$ and $\kappa=1.7$. We
conclude that when $\kappa=1.7$, the contribution from a T-odd fermion to the $T$
parameter is close to the maximal value only for small $f$, but when
$\kappa=0.7$, the contribution from a T-odd fermion is about one order of magnitude
below. In the latter case we will assume that these kind of contributions to the $T$
parameter can be neglected. 

\subsection{Scenario with heavy T-odd fermions}
\begin{figure}[!ht]
 \centering
\includegraphics[width=3.5in]{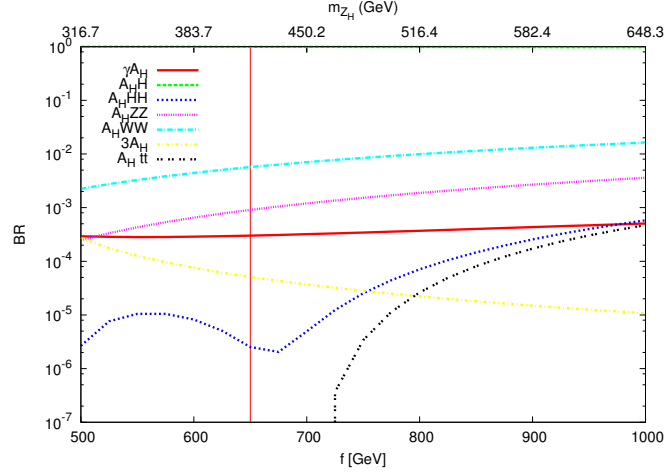}
 \caption{\label{BRfLHT1} Branching ratio for the one-loop decay $Z_H\to \gamma  A_H$  in
the LHTM as a function of the scale of symmetry breaking $f$. We also include the main
tree-level two- and three-body decays. We used  $m_H=120$ GeV,
$s_\lambda=0.55$, and $\kappa=1.7$. The region to the right of
the vertical line is not allowed by EWPD.}
 \end{figure}

According to the above discussion, we will consider the following values for the LHTM
parameters: $m_H=120$ GeV, $s_\lambda=0.55$, and $\kappa=1.7$.
In Fig. \ref{BRfLHT1} we show the branching ratio for
the $Z_H\to\gamma A_H$ decay together with
those of the relevant tree-level decays as functions of the scale of the symmetry
breaking.   The Loop-Tools package
\cite{vanOldenborgh:1989wn,Hahn:2000jm} was used to evaluate the Passarino-Veltman scalar
functions required by the $Z_H\to \gamma A_H$ decay width. In this scenario, the dominant
decay mode is $Z_H\to A_H H$, which has a
branching ratio of about $100\%$ for $f\simeq 500$ GeV
and about $80\%$ for $f\simeq 2$ TeV. Other  kinematically
allowed tree-level decays, such as $Z_H\to A_H HH$, $Z_H\to A_H WW$, $Z_H\to A_H ZZ$,
$Z_H\to A_H \bar{t}t$ and $Z_H\to A_H A_H A_H$, have  branching ratios of the order of
$1\%$ for $f\simeq 500$ GeV, which increase as $f$ increases. In particular, the $Z_H\to
A_H WW$ branching ratio increases up to $10\%$ for $f\simeq 2$ TeV. Although  decays
into a heavy photon plus a pair of light fermions are also kinematically allowed,  their
decay width is negligibly as the main contribution arises from a Feynman diagram where the
fermion pair is emitted off the Higgs boson. We
also observe that the branching ratio for the decay into three heavy photons can be of
similar size
than the ones for other three-body decays in the region of low $f$ but it decreases
quickly as $f$
increases. This decay proceeds as follows: for $f$ below about $465$ GeV, the $m_{A_H}$ is
below 60 GeV and thus the $Z_H$ gauge boson decays into a heavy photon plus a real Higgs
boson with a mass of $120$ GeV, which subsequently decays into a heavy photon pair. For
$f>465$ GeV, $m_{A_H}>m_H/2$, so the intermediate Higgs boson is virtual. As for the
one-loop decay $Z_H\to \gamma A_H$, the respective branching ratio is of the order of less
than one percent for $f=500$ GeV, but it increases slightly for $f=2$ TeV. We would
like to note that the $Z_H$ branching ratios are not very sensitive to a change in
the value of the $s_\lambda$ parameter.

\begin{figure}[!ht]
\centering
\includegraphics[width=3.5in]{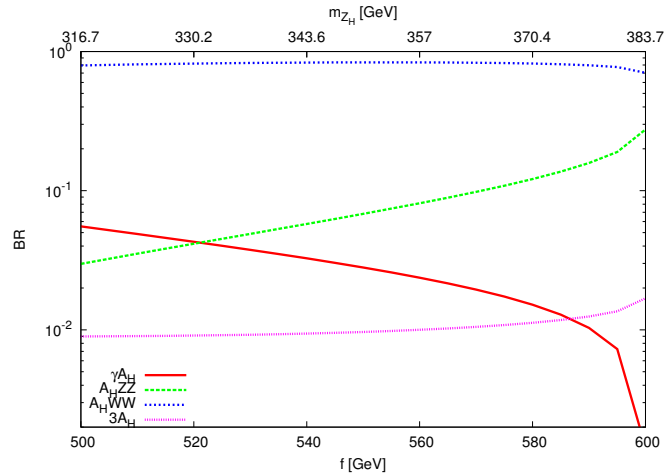}
\caption{\label{BRfLHT2} The same as in Fig. \ref{BRfLHT1}, but for $m_H=300$ GeV.}
 \end{figure}

An interesting scenario arises  when the Higgs boson has an intermediate mass such that
$m_H> Z_H-A_H$. In this case the $Z_H\to A_H H$ decay channel will be closed and the
$Z_H\to
\gamma A_H$ decay has a substantial  enhancement. This situation occurs, for
instance, for $m_H$ about 300 GeV and $f$ up to 600 GeV or either for $m_H$ about 500 GeV
and $f$ up to 1 TeV. Such scenarios are  still
allowed by EWPD for some particular values of $s_\lambda$. We have
calculated the branching ratios of the main $Z_H$ decays for $m_H=300$ GeV,
$s_\lambda=0.55$ and $\kappa=1.7$. We show the results in Fig. \ref{BRfLHT2}. The dominant
decay mode is now $Z_H\to A_HWW$, whereas the $Z_H\to \gamma A_H$ decay width is even
larger than the one for the $Z_H\to A_HZZ$ decay. In general, $BR(Z_H\to \gamma
A_H)$ gets enhanced up to two orders of magnitude with respect to what is
obtained in the scenario with a light Higgs boson. When $f\simeq 600$ GeV,  $BR(Z_H\to
\gamma A_H)$ drops suddenly as  the threshold for the opening of the $Z_H\to
A_H H$ decay is reached.

\subsection{Scenario with relatively light T-odd fermions}
\begin{figure}[!ht]
 \centering
\includegraphics[width=3.5in]{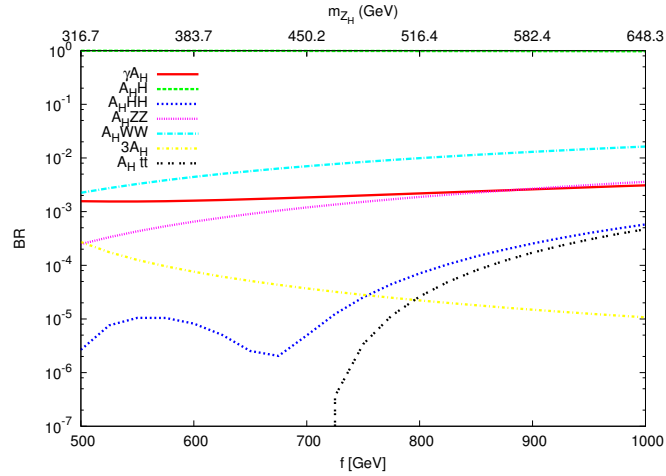}
 \caption{\label{BRfLHT3} The same as in Fig. \ref{BRfLHT1}, but for $s_\lambda=0.75$, and
$\kappa=0.7$.}
 \end{figure}

\begin{figure}[!ht]
 \centering
\includegraphics[width=3.5in]{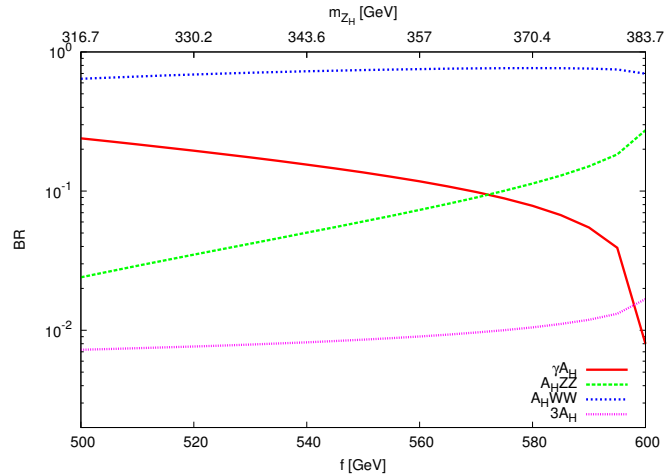}
 \caption{\label{BRfLHT4} The same as in Fig. \ref{BRfLHT3}, but for $m_H=300$ GeV.}
 \end{figure}

We now consider the case of  $m_H=120$
GeV, $s_\lambda=0.75$ and
$\kappa=0.7$. This means that the T-odd fermions are relatively light and their
contribution to the $T$ parameter is small. We show in Fig. \ref{BRfLHT3} the relevant
$Z_H$ branching ratios for this scenario. We observe that the
$Z_H\to \gamma A_H$ branching ratio is enhanced by about one order of magnitude with
respect to the  scenario with heavy T-odd fermions. Even more, the $Z_H\to \gamma A_H$
branching ratio can be as large as the $Z_H\to A_H WW$ branching ratio for small $f$, but
the latter increases steadily with $f$ and it becomes much larger for $f$ close to 1 TeV.
Overall the $Z_H\to \gamma A_H$ branching ratio is of the same order
of magnitude as the one for the $Z_H\to A_H ZZ$ decay over a large range of $f$. We also
would like to consider the
case of an intermediate Higgs boson. We show  in Fig.
\ref{BRfLHT4} the $Z_H$ branching ratios for 
$s_\lambda=0.75$, $\kappa=0.7$, and $m_H=300$ GeV.
The situation looks similar to that depicted in Fig. \ref{BRfLHT2}. The branching
ratio for the $Z_H\to \gamma A_H$ decay shows a large increase and it is the subdominant
$Z_H$ decay channel for up to $f\simeq 570$ GeV, where it starts to decrease quickly.
Although the $Z_H\to \gamma A_H$ decay can have a substantial enhancement over a wide $f$
range if the Higgs boson is heavier, the $Z_H$ production cross section decreases quickly
as $f$ increases, so the $Z_H\to \gamma A_H$ decay would be more difficult to detect for
$f$ close to $1$ TeV.

\subsection{Experimental prospects at the LHC}
\begin{figure}[!ht]
 \centering
\includegraphics[width=3.5in]{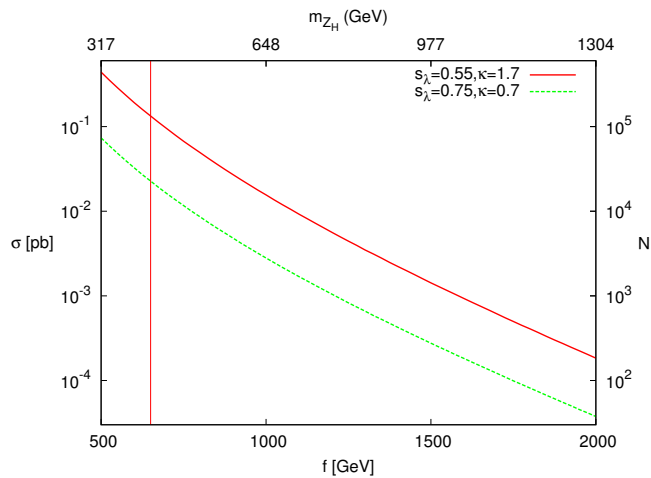}
\caption{\label{seceflht1} Cross section for $pp\to W_HZ_H$ at the LHC in the two
scenarios described in the text. In this plot $\sqrt{s}=14$ TeV and the CTEQ6M PDF
set was used. The expected
number of events is shown in the left axis for a luminosity of 300 fb$^{-1}$. The region
to the right of
the vertical line is not allowed for $s_\lambda=0.55$ by EWPD.}
 \end{figure}

At the LHC, the T-odd gauge bosons must
be pair produced due to T-parity invariance. The dominant $Z_H$
production mode is $pp\to W_HZ_H$, whereas other production modes such as $pp\to Z_H Z_H$
and $pp\to A_H Z_H$ are suppressed by more than one and two orders of magnitude,
respectively.
We show in Fig. \ref{seceflht1} the dominant
$Z_H$ production mode at the LHC for the two scenarios described above and $\sqrt{s}=14$
TeV. The CTEQ6M PDF set
was used.\cite{Pumplin:2002vw} This calculation was obtained via the CalcHep package
\cite{Pukhov:2004ca} along with the LHTM files provided  by the authors of Ref.
\refcite{Belyaev:2006jh}.  In Fig. \ref{seceflht1} we also observe that  a luminosity of
$300$ fb$^{-1}$ would allow for a large number of $pp\to W_H Z_H$ events, of the
order of  $10^5$ for $f=500$ GeV and $10^2$ for $f=2$ TeV.

\begin{figure}[!ht]
 \centering
\includegraphics[width=3.5in]{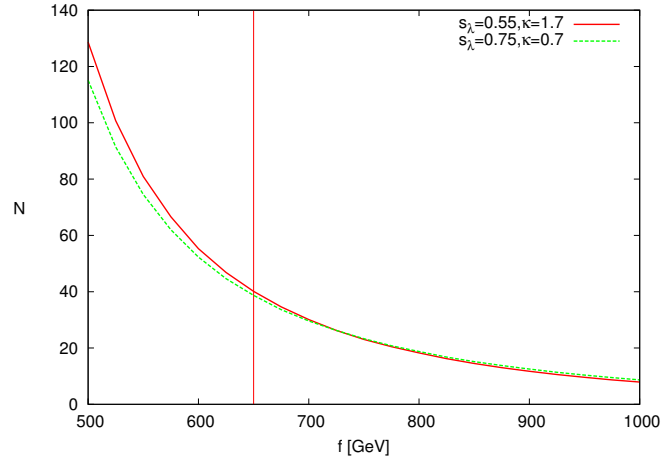}
 \caption{\label{neven13} Expected number of $pp\to W_H Z_H\to A_H W +\gamma A_H$ events
at the LHC for $\sqrt{s}=14$ TeV in the scenario with a light Higgs boson with a mass
$m_H=120$ GeV. A luminosity of 300 fb$^{-1}$ was considered. The region to the right of
the vertical line is not allowed for $s_\lambda=0.55$ by EWPD.}
 \end{figure}

\begin{figure}[!ht]
 \centering
\includegraphics[width=3.5in]{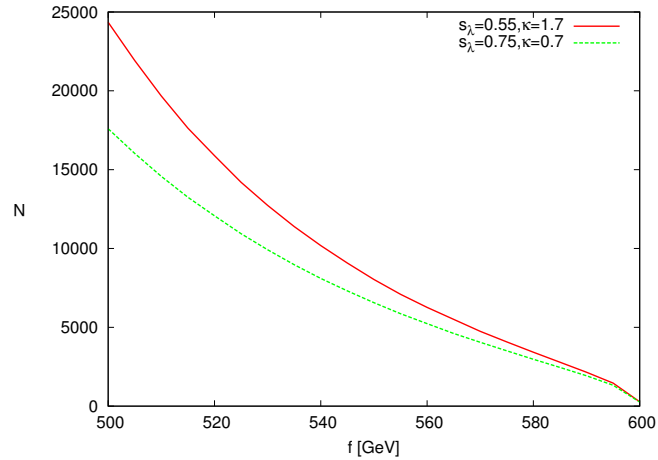}
 \caption{\label{neven14} The same as in Fig. \ref{neven13}, but for $m_H=300$ GeV.}
 \end{figure}

Assuming that the $W_H$
gauge boson decays as $W_H\to A_HW$ with a rate of $100\%$, we have calculated the
expected number of $pp\to W_HZ_H\to W A_H +\gamma A_H$ events for a luminosity of 300
fb$^{-1}$. For the scenarios
discussed above, the dependence on the scale of the symmetry breaking of the expected
number of $pp\to W_HZ_H\to W A_H +\gamma A_H$ events is
shown in Fig. \ref{neven13} (light Higgs boson) and Fig. \ref{neven14} (intermediate
Higgs boson). In the case of a light Higgs boson, we
observe that the event number is of the order of one  hundred around $f=500$ GeV and
decreases quickly as $f$ increases. As far as the scenario with an intermediate Higgs
boson is concerned, there would be a large number of events for $f=500$ GeV, but a sharp
decrease is observed for $f$ about $600$ GeV, where the $Z_H\to A_H H$ channel gets
opened.

The experimental signature of the $Z_H\to \gamma A_H$ decay would be a charged lepton
accompanied by an energetic photon plus large missing transverse energy \met. The main
background comes from the SM process
$pp\to W\gamma\to \ell\nu\gamma$, but it can be largely reduced by imposing cuts on the
photon energy. In the scenario with a light Higgs boson, the $Z_H\to\gamma A_H$ detection
seems promising, but a more detailed Monte
Carlo analysis would be required to make further conclusions.  As for the scenario with
an intermediate Higgs boson is concerned, although there is a large number of $WA_H+\gamma
A_H$ events for $f\simeq 500$ GeV, it drops drastically for $f\simeq 600$ GeV, which would
make the $Z_H \to \gamma A_H$ decay detection very difficult for large $f$.

\section{Final remarks}
We have examined the one-loop decay $Z_H\to \gamma A_H$ in the framework of the LHTM.
Since T-parity restricts the number of possible decay channels of the heavy $Z_H$ gauge
boson, one-loop $Z_H$ decays can have branching fractions of similar order of magnitude
as some $Z_H$ tree-level three-body decays. One advantage of the $Z_H\to \gamma A_H$ decay
channel is that it
would not require the previous detection of the Higgs boson. We have examined two
potential scenarios still allowed by EWPD: relatively light or heavy T-odd fermions. In
the case of a light Higgs boson, with a
mass of 120 GeV, the expected
number of $pp\to W_H Z_H \to A_HW+ \gamma A_H$ events is of the order of one hundred for
$f\simeq 500$ GeV, but it decreases quickly for increasing $f$. The possible detection of
the $Z_H\to \gamma A_H$ decay looks thus only favorable for $f$ about $500$ GeV.
Other $Z_H$ production modes, such as $pp\to A_H Z_H$ and $pp\to Z_H Z_H$, would give less
than 10 $Z_H\to \gamma A_H$ events for $f\simeq 500$ GeV and would not be useful to detect
this decay mode. An interesting scenario is that in which there is an intermediate Higgs boson with a mass of
the order of $300$ GeV. From 500 GeV to 600 GeV, the $Z_H\to A_H H$ is closed and the
event number of $pp\to W_H Z_H \to A_HW+ \gamma A_H$ events gets enhanced by more than
three orders of magnitude with respect to the case of a light Higgs boson. This scenario
would be more favorable for the detection of the $Z_H\to \gamma A_H$ decay but also requires that $f$ is not too large to allow the opening of the $Z_H\to A_H H$ channel.

\section*{Acknowledgments}GTV acknowledge support from Conacyt and SNI (M\'exico). We also acknowledge support from VIEP-BUAP.

\appendix
\section{Form factors for the $Z_H\to \gamma A_H$ decay}
\label{Coefficients}
In this appendix we present the explicit form for the $A_i^{\gamma A_H}$ coefficients in
terms of Passarino-Veltman scalar functions. They are given as
follows:

\begin{eqnarray}
 A_1^{\gamma A_H}&=&\left(\frac{g}{8\pi c_W}\right)^2 \frac{2}{\left(1 - y_{A_H}\right)^3
y_{A_H}}\sum_{f_+}\xi^f_{\gamma A_H} \Big\{ y_{A_H}^3
\Big[2+B_{c_\pm}-B_{a_+}\nonumber\\&+&2((2 y_{f_+}+1)C_{a_\mp}-
y_{f_-}(C_{a_\mp}-C_{a_\pm}))\Big]\nonumber\\&+& y_{A_H}^2
\Big[(y_{f_+}-y_{f_-})(3B_{c_\pm}-2 B_{a_+}-B_{a_\pm})-2(1+ 3
(B_{b_\pm}-B_{c_\pm}))\nonumber\\&-&2(2  y_{f_-}
C_{a_\pm}+(1-(y_{f_+}-y_{f_-})^2)C_{a_\mp})\Big]\nonumber\\
&+&y_{A_H} \Big[2(y_{f_+}-y_{f_-})(2B_{c_\pm}+B_{a_+}-3B_{b_\pm})+
B_{a_+}-B_{c_\pm}\nonumber\\ &-&2(((y_{f_+}-y_{f_-})^2+2y_{f_+}-y_{f_-}) C_{a_\mp}-
y_{f_-} C_{a_\pm})\Big]\nonumber\\&+&(y_{f_+}-y_{f_-})(B_{a_\pm}-B_{c_\pm})\Big\},
\end{eqnarray}

\begin{eqnarray}
 A_2^{\gamma A_H}&=&\left(\frac{g}{8\pi c_W}\right)^2 \frac{2}{\left(1 - y_{A_H}\right)^3
y_{A_H}}\sum_{f_+}\Big\{\xi^f_{\gamma A_H} \Big[ y_{A_H}^3
\Big(B_{a_+}-B_{c_\pm}\nonumber\\&-&2(1+ (y_{f_+}-y_{f_-}+1)C_{a_\mp})\Big)+ y_{A_H}^2
\Big((y_{f_+}-y_{f_-})(2B_{a_+}+B_{a_\pm}-3B_{c_\pm})\nonumber\\&+&2(1+B_{b_\pm}-B_{c_\pm}
)+2(2y_{f_-} C_{a_\pm}-(y_{f_+}-y_{f_-})^2 C_{a_\mp}+C_{a_\pm})\Big)\nonumber\\
&+&y_{A_H}
\Big(B_{c_\pm}-B_{a_+}+2(y_{f_+}-y_{f_-})((3B_{b_\pm}-2B_{c_\pm}-B_{a_+})\nonumber\\&+&(y_
{f_+}-y_{f_-}+1)C_{a_\mp})-4y_{f_-}C_{a_\pm}\Big)+(y_{f_+}-y_{f_-})(B_{c_\pm}-B_{a_\pm}
)\Big]\nonumber\\&+&2\lambda^f_{\gamma A_H}(1-y_{A_H})^2\sqrt{y_{f_+}}\sqrt{y_{f_-}}
(C_{a_\pm}+C_{a_\mp})\Big\},\nonumber\\
\end{eqnarray}

\begin{eqnarray}
 A_3^{\gamma A_H}&=&\left(\frac{g}{8\pi c_W}\right)^2 \frac{2}{\left(1 -
y_{A_H}\right)^3}\sum_{f_+}\Bigg\{\tilde{\xi}^f_{\gamma A_H} \bigg[
2y_{A_H}^2\left(y_{f_-}C_{a_\pm}-y_{f_+}C_{a_\mp}\right)\nonumber\\&+&
\frac{4}{3}y_{A_H}\big(y_{f_+}(2B_{a_+}+B_{a_\pm}-3B_{c_\pm})
-y_{f_-}(2B_{a_-}+B_{a_\pm}-3B_{c_\pm})\nonumber\\&-&(y_{f_+}-y_{f_-})(1+3(y_{f_-}C_{a_\pm
}+y_{f_+}C_{a_\mp}))\big) \nonumber\\
&+&\frac{2}{3}\Big(2 y_{f_-} (B_{a_\pm} +2B_{a_-}+3 B_{c_\pm}-6
B_{b_\pm}-1)\nonumber\\&-&2 y_{f_+} (2 B_{a_+} + B_{a_\pm} + 3 B_{c_\pm} - 6 B_{b_\pm} -
1)\nonumber\\
&+&6(y_{f_+}-y_{f_-}) (y_{f_-}C_{a_\pm}-y_{f_+}C_{a_\mp})
+3(y_{f_+}C_{a_\mp}-y_{f_-}C_{a_\pm})\Big) \bigg]\nonumber\\&+&2\tilde{\lambda}^f_{\gamma
A_H}(1-y_{A_H})^2\sqrt{y_{f_+}}\sqrt{y_{f_-}} (C_{a_\mp}-C_{a_\pm})\Bigg\},\\
A_4^{\gamma A_H}&=&\frac{1}{y_{A_H}} A_3^{\gamma A_H}\left(y_{A_H}\leftrightarrow
\frac{1}{y_{A_H}}\right),
\end{eqnarray}
where the scalar functions are as follows

\begin{eqnarray}
B_{a_+}&=&B_0(0,
m_{f_+}^2, m_{f_+}^2),\\
B_{a_\pm}&=& B_0(0, m_{f_+}^2, m_{f_{-}}^2),\\
B_{a_-}&=& B_0(0, m_{f_{-}}^2, m_{f_{-}}^2),\\
B_{b_\pm}&=& B_0(m_{Z_H}^2, m_{f_+}^2,m_{f_{-}}^2),\\
B_{c_\pm}&=& B_0(m_{A_H}^2, m_{f_+}^2,m_{f_{-}}^2),\\
C_{a_\pm}&=&m_{Z_H}^2C_0(m_{A_H}^2, 0, m_{Z_H}^2, m_{f_+}^2, m_{f_{-}}^2, m_{f_{-}}^2),\\
C_{a_\mp}&=&m_{Z_H}^2C_0(m_{A_H}^2, 0, m_{Z_H}^2, m_{f_{-}}^2, m_{f_+}^2, m_{f_+}^2).
\end{eqnarray}
Also
\begin{eqnarray}
\xi^f_{\gamma A_H}&=& N_c^f Q^f  \left({g'_L}^f {g''_L}^f+{g'_R}^f
{g''_R}^f\right),\\
\tilde{\xi}^f_{\gamma A_H}&=& N_c^f Q^f  \left({g'_L}^f {g''_L}^f-{g'_R}^f
{g''_R}^f\right),\\
\lambda^f_{\gamma Z}&=&N_c^f Q^f  \left({g'_L}^f
{g''_R}^f+ {g'_R}^f {g''_L}^f\right),\\
\tilde{\lambda}^f_{\gamma Z}&=&N_c^f Q^f  \left({g'_L}^f
{g''_R}^f- {g'_R}^f {g''_L}^f\right),
\end{eqnarray}
where  ${g'_{L,R}}^f$ (${g''_{L,R}}^f$) are the
constants associated with the $Z_H\bar{f}_+f_-$
($A_H\bar{f}_+f_-$) coupling (See Appendix A).  The sum runs over each charged T-even
fermion and
its associated
T-odd fermion. In the case of the top sector, the fermions to be summed are the pairs $(t,
t_-)$,
$(t,T_-)$, $(T_+,t_-)$, and $(T_+,T_-)$.
Notice that $A_3^{\gamma A_H}$ and $A_4^{\gamma A_H}$ are antisymmetric under the exchange
 $f_+
\leftrightarrow f_-$, and thus vanish when the same fermion circulates into the loop. In
addition,
the transition amplitude vanishes when $y_{A_H}=1$ as required for the on-shell
$Z_HZ_H\gamma$
vertex.

\section{Feynman rules for the $Z_H$ and $A_H$ gauge bosons in the LHTM}
\label{Couplings}
In this appendix we collect all the Feynman rules necessary for our calculation. They were taken
from Refs.  \refcite{Belyaev:2006jh} and \refcite{Han:2003wu}.

\subsection{Couplings to T-even and T-odd fermions}

We write the couplings of the extra neutral $Z_H$ gauge boson to T-even and  T-odd fermions in
the form:
\begin{equation}
{\cal L}=  -\frac{ig}{c_W }\bar{f_+}\gamma_{\mu}\left({g'}^f_L P_L+{g'}^f_R
P_R\right)f_-{Z_H}^\mu,
\label{ZHffLag}
\end{equation}
with $P_{L,R}$ the usual chirality projectors. A similar expression holds for
the $A_H$ gauge boson couplings with the replacement ${g'}^f_{L,R}\to {g''}^f_{L,R}$. The respective coupling constants are
shown in Table \ref{LHTMVfefo}.

\begin{table}[!ht]
\tbl{Couplings of the the heavy neutral gauge bosons to a  T-even antifermion and a T-odd
fermion in the LHTM.  The second line is valid for all the down-type fermions, but the
first line is valid for all up-type fermions other than those of the top sector. The
mixing angle is $s_H\sim gg'v^2/(g^2-g'^2/5)/(4f^2)$, with
$c_H^2=1-s_H^2$.}
{\begin{tabular}{|c|c|c|c|c|}
\hline
&\multicolumn{2}{|c|}{$V=Z_H$}&\multicolumn{2}{|c|}{$V=A_H$}\\
\cline{2-5}
\hline
 & ${g'_L}^f$ &${g'_R}^f$ & ${g''_L}^f$ &${g''_R}^f$   \\
\hline
$V \bar{u} u_-$  & $\frac{g c_H}{2}-\frac{g's_H}{10}$ &$0$& $-\frac{g
s_H}{2}-\frac{g'c_H}{10}$
&$0$\\
\hline
$V \bar{d} d_-$ & $-\frac{g c_H}{2}-\frac{g's_H}{10}$ & 0&$\frac{g
s_H}{2}-\frac{g'c_H}{10}$ &$0$\\
\hline
$V \bar{t} t_-$ & $\left(\frac{g c_H}{2}-\frac{g's_H}{10}\right)c_L$&0&$\left(-\frac{g
s_H}{2}-\frac{g'c_H}{10}\right)c_L$ &$0$\\
\hline
$V \bar{t} T_-$ &
$-\frac{2g's_H}{5}s_L$&$-\frac{2g's_H}{5}s_R$&$-\frac{2g'c_H}{5}s_L$&$-\frac{2g'c_H}{5}
s_R$\\
\hline
$V \bar{T_+} t_-$ & $\left(\frac{g c_H}{2}-\frac{g's_H}{10}\right)s_L$&$0$&$\left(-\frac{g
s_H}{2}-\frac{g'c_H}{10}\right)s_L$&\\
\hline
$V \bar{T_+} T_-$
&$\frac{2g's_H}{5}c_L$&$\frac{2g's_H}{5}c_R$&$\frac{2g'c_H}{5}c_L$&$\frac{2g'c_H}{5}c_R$\\
\hline
\end{tabular}\label{LHTMVfefo}}
\end{table}

\subsection{Couplings involving two heavy gauge bosons}

We also need Feynman rules for the couplings  $Z_H A_H H$, $A_H A_H H$,  and $Z_H A_H HH$.
These vertices are involved in the tree-level decays of the $Z_H$ gauge boson. The
respective Feynman rules are shown in Table \ref{AHcoup} together with the Feynman rules
for the trilinear and quartic gauge boson couplings involved in our calculation. We have
defined

\begin{equation}
F^{\mu\nu\rho}(k_1,k_2,k_3)= g^{\mu\nu} (k_1 - k_2)^{\rho}+
g^{\nu\rho} (k_2 - k_3)^{\mu} + g^{\rho\mu} (k_3 - k_1)^{\nu},
\label{trilinear}
\end{equation}
where all particles are outgoing, and
\begin{equation}
G^{\mu\nu\rho\sigma}= 2g^{\mu\nu}g^{\rho\sigma}-g^{\mu
\rho}g^{\nu\sigma}-g^{\mu\sigma}g^{\nu\rho}.
\label{quartic}
\end{equation}

\begin{table}[!ht]
\tbl{Feynman rules for the vertices involved in the calculation of the $Z_H$ gauge boson decays in the LHTM.}
{\begin{tabular}{|c|c|}
\hline
 Vertex& Feynman rule  \\
\hline
${A_H^\mu A_H^\nu H}$&$-i\frac{g' v^2}{2}g^{\mu\nu}$\\
\hline
${Z_H^\mu A_H^\nu H}$&$-i\frac{g g' v }{2}g^{\mu\nu}$\\
\hline
${Z_H^\mu A_H^\nu HH}$&$-i\frac{g g' }{2}g^{\mu\nu}$\\
\hline
${A_H^\mu(k_1) W_H^\nu(k_2) W^\rho(k_3)}$&$i\frac{5g}{4(5-t_W^2)}\frac{v^2}{f^2}F^{\mu\nu\rho}(k_1,k_2,k_3)$\\
\hline
${Z_H^\mu(k_1) W_H^\nu(k_2) W^\rho(k_3)}$&$ig F^{\mu\nu\rho}(k_1,k_2,k_3)$\\
\hline
 ${ W^\mu W^\nu Z_H^\rho A_H^\sigma}$&$- i\frac{5 g^2 v^2}{4(5-t_W^2)f^2} G^{\mu\nu\rho\sigma}$\\
\hline
\end{tabular}\label{AHcoup}}
\end{table}

\end{document}